# Effective permittivity of random particulate plasmonic composites

**Satvik N. Wani,[1] Ashok S. Sangani,[1,2] and Radhakrishna Sureshkumar[1,3,*]**

[1]*Department of Biomedical and Chemical Engineering, Syracuse University, Syracuse, NY 13244, USA*

[2]*National Science Foundation, Arlington, VA 22230, USA (On leave from Syracuse University)*

[3]*Department of Physics, Syracuse University, Syracuse, NY 13244, USA*

**Corresponding author: rsureshk@syr.edu*

An effective-medium theory (EMT) is developed to predict the effective permittivity $e_{eff}$ of dense random dispersions of high optical-conductivity metals such as Ag, Au and Cu. Dependence of $e_{eff}$ on the volume fraction $f$, a microstructure parameter $k$ related to the static structure factor and particle radius $a$ is studied. In the electrostatic limit, the upper and lower bounds of $k$ correspond to Maxwell-Garnett and Bruggeman EMTs respectively. Finite size effects are significant when $|b^2(ka/n)^3|$ becomes $O(1)$ where $b$, $k$, and $n$ denote the nanoparticle polarizability, wavenumber and matrix refractive index respectively. The coupling between the particle and effective medium results in a red-shift in the resonance peak, a non-linear dependence of $e_{eff}$ on $f$, and Fano resonance in $e_{eff}$.

*OCIS codes: 160.4236, 160.4670, 260.2065, 260.2110, 260.3910, 260.5740.*

## 1. Introduction

Nanoparticulate plasmonic composite materials have recently become technologically important, especially in the growing interdisciplinary fields of plasmonics and meta-materials [1-6]. Fabrication of such plasmonic nanocomposites is accomplished through well established methods such as surfactant mediated self-assembly [7], laser dewetting of thin films [8], sol-gel assembly [9-11], ion-implantation [12-13] and vacuum evaporation of thin films [14] among others. Technologically, such composites are important in the fields of energy harvesting [7, 15-17], random lasers [18], sensing [16, 19] and photo-catalysis [20] etc. The free electrons in noble metal nanoparticles (NPs) give rise to a characteristic plasmon resonance wherein the NPs absorb and scatter radiation with a marked intensity [3, 21-23]. The linear optical response of such materials can be described by Maxwell equations in both the dispersed and continuous phases when the particle size is greater than a few nanometers. In the case of metal nanoparticles in a transparent medium, the matrix phase can be treated as a dielectric with real electrical permittivity $\varepsilon_m$ while the NPs should be treated as materials with complex, frequency-dependent permittivity $\varepsilon_p$. A plasmon resonance occurs for $\Re(\varepsilon_p/\varepsilon_m) = -2$ in the case of spheres. The overall optical response of such composite materials can be determined by numerically solving Maxwell equations in both phases subject to the continuity of tangential components of magnetic and electric fields at the interface of the embedded particles and the surrounding matrix phase. Such calculations are possible by utilizing computational techniques such as the finite difference time domain method [24-26]. However, substantial computational effort is required especially near resonant frequencies where steep field gradients necessitate the use of very fine spatial resolution [26]. Hence, theories capable of accurately predicting the average optical properties of random composites could be a valuable tool in knowledge-based design of plasmonic composites.

Composite media are inherently inhomogeneous. Hence, their average electromagnetic behavior depends on the permittivities and volume fractions of the constituent components. Effective medium theories have been used to parameterize the properties of such media [27]. In the case of monodisperse spherical particulate composites, where the particle radius $a$ is much smaller than the wavelength of exciting radiation $\lambda$, the effective permittivity can be modeled under the quasistatic approximation wherein the wave nature of the EM fields can be neglected. Specifically, for the case of plasmonic composites in the optical frequency range, the magnetic response in the optical range is the same as that of vacuum and the classical Maxwell-Garnett theory (MGT) [28] can be used to predict the effective permittivity. Similarly, in the case $\lambda << a$, ray optics can be utilized. However, for nanoscopic plasmonic composites in the optical range, $\lambda$ can be $O(a)$ or $ka \equiv 2\pi\sqrt{\varepsilon}a/\lambda$ is $O(1)$. Hence, the effects of diffraction and scattering by the NPs become significant and simple models designed for either one of the extreme cases are not applicable. Extended Maxwell-Garnett theories have been developed in the literature for this regime [29-32]. However they are correct only up to $O(f)$ ($f$: volume fraction of the dispersed phase) and do not account for the effect of microstructure on the permittivity. In this work, we have developed a self-consistent theoretical framework for the prediction of the effective linear optical properties of dense random monodisperse spherical particulate plasmonic composites with particle size on the order of the exciting wavelength of radiation. This effective-medium theory (EMT) is based a method employed for the accurate prediction of sound attenuation and phase speed in acoustically resonant monodisperse suspensions of microspheres by Spelt et al. [33]. The microstructure information is incorporated through the static structure factor $S(\mathbf{0})$. It has been shown that the leading order correction term in terms of the particle volume fraction to the velocity field in Stokes flow, i.e., "slow" flow of a viscous liquid in which inertial forces are negligibly small, for infinite randomly distributed monodisperse spheres depends linearly on the static structure factor $S(\mathbf{0})$ [34]. Both Stokes flow equations and Maxwell wave equations, take the form of a vector Helmholtz equation for the fluid velocity and electric field respectively. Hence, the methodology developed by Spelt et al. [17] can be adapted to derive an EMT for electromagnetic wave propagation problems in heterogeneous media. Specifically, the composite medium is represented by a layered structure in which the particle, in its immediate vicinity, is surrounded by the dielectric matrix up to a distance $R$ which is a function of $S(\mathbf{0})$. The structure is assumed to be embedded in an effective continuum whose permittivity $\epsilon_{eff}$ is determined by the self-consistent solution of Maxwell equations. The EMT is mathematically identical to the Maxwell-Garnett model in the limit as the particle diameter $d_p$ and volume fraction $f$ approach zero. However, for finitely large volume fractions the variations in the permittivity with respect to $f$ and $d_p$ are highly nonlinear.

The paper is organized as follows. Problem formulation is presented in §2. §2A contains the derivation of the ensemble averaged Maxwell equations and in §2B, the EMT is discussed. §3 contains a summary of the solution technique and computational methods used to calculate the conditionally averaged electric field. Analytical and numerical results are discussed in §4. Results of the scalar EMT is discussed in §4A and those of vector EMT in §4B. Ag nanoparticulate composite in an $\epsilon_m = 7$ dielectric is used as a model system. A discussion of the conditions under which the effective permittivity is resonant is presented in subsections §§4A and B. Effects of particle radius are also discussed in §4B. §4C contains a discussion of Fano-resonance that results from particle-effective medium coupling. Conclusions are offered in §5.

## 2. Ensemble averaged Maxwell equation

We consider a random monodisperse, non-overlapping spherical particulate composite in which the electrical permittivity of the matrix is assumed to be real, positive and constant while that of the plasmonic particles is complex and frequency-dependent. Further, as mentioned in the Introduction, we assume that the magnetic permeabilities of the matrix and the particle phase to be equal to that of vacuum, i.e., $\mu_m = \mu_p = \mu_0$. This assumption is justified for dielectric matrices such as glass or water and particles of noble metals such as Ag or Au. The particle diameter is assumed to be much greater than the electron mean free path in the metal. Hence, quantum confinement effects are neglected. The embedding medium is isotropic and homogenous and could either be a liquid or a solid phase. For such a system, ensemble-averaged Maxwell equations can be derived as described below.

Time-harmonic electric and magnetic fields in a source-free homogenous medium satisfy Maxwell's wave equations. For the matrix these equations are given by

$$\nabla \times \mathbf{E}_m = i\omega\mu_m \mathbf{H}_m, \quad \nabla \times \mathbf{H}_m = -i\omega\varepsilon_m \mathbf{E}_m, \tag{1}$$

where $\mathbf{E}_m$ and $\mathbf{H}_m$ are the amplitudes of the electric and magnetic fields respectively, $\omega$ is the frequency, $\mu_m$ and $\varepsilon_m$ are the magnetic permeability and electric permittivity respectively, and the subscript $m$ denotes the matrix medium. Similar equations apply in the particulate phase with the subscript $m$ replaced by the particle phase subscript $p$.

To obtain a macroscopic description of a random composite, we must first obtain ensemble-averaged equations. Let $g_p$ denote an indicator function for the particle phase whose value at a point $\mathbf{x}$ is unity if that point lies inside a particle and zero otherwise. Note that an ensemble-average of this function is equal to the volume fraction of the particles, i.e., $\langle g_p \rangle(\mathbf{x}) = f$, where the angular brackets denote an ensemble-averaged quantity. The ensemble-averaged Maxwell's equations for the random composites are obtained by multiplying the Maxwell's equations for the particle phase by its indicator function $g_p$ and those for the matrix phase by $1 - g_p$, and adding the two resulting equations:

$$\nabla \times \langle \mathbf{E} \rangle + \langle \nabla g_p \times (\mathbf{E}_m - \mathbf{E}_p) \rangle = i\omega\mu_0 \langle \mathbf{H} \rangle, \tag{2a}$$

$$\nabla \times \langle \mathbf{H} \rangle + \langle \nabla g_p \times (\mathbf{H}_m - \mathbf{H}_p) \rangle = -i\omega \left[ \varepsilon_m \langle \mathbf{E} \rangle + (\varepsilon_p - \varepsilon_m) \langle g_p \mathbf{E} \rangle \right]. \tag{2b}$$

Note that $\nabla g_p$ is zero at all points except at the matrix-particle interface where it is directed along the normal to the interface. Hence, its cross product with the difference in **E** or **H** across the interface is zero due to the fact that the tangential components of the electric and magnetic fields are continuous across the interface. Therefore, the second terms on the left-hand side of Eqs. 2a and 2b vanish. The term in square brackets on the right-hand side of Eq. 2b is represents the averaged electric displacement $\langle \mathbf{D} \rangle$ in the medium. We let $\langle \mathbf{D} \rangle = \varepsilon_{eff} \langle \mathbf{E} \rangle$. Hence, the effective permittivity can be defined as

$$\varepsilon_{eff} \langle \mathbf{E} \rangle = \varepsilon_m \langle \mathbf{E} \rangle + (\varepsilon_p - \varepsilon_m) \langle g_p \mathbf{E} \rangle. \tag{3}$$

The wavenumber, defined as $k_n^2 = \omega^2 \mu_n \varepsilon_n, n = p, m$, also obeys Eq. 3. Hence, the effective wavenumber is given by $k_{eff}^2 = \omega^2 \mu_0 \varepsilon_{eff}$. Eqs. 2a, 2b and 3 can be combined using a curl operation on Eq. 2a. The resulting $\nabla \times \nabla \times \langle \mathbf{E} \rangle$ term can be shown to be equal to $-\nabla^2 \langle \mathbf{E} \rangle$ because $\nabla \times (\nabla \langle \mathbf{E} \rangle) = \mathbf{0}$ in the absence of free charge.

The averaged field inside the particles, given by $\langle g_p \mathbf{E} \rangle$, is an unknown quantity defined as:

$$\langle g_p \mathbf{E} \rangle(\mathbf{r}) \equiv \int_{|\mathbf{r}-\mathbf{r}_1| \le a} \langle \mathbf{E}(\mathbf{r}) \,|\, \mathbf{r}_1 \rangle P(\mathbf{r}_1)\, dV_{\mathbf{r}_1}. \tag{4}$$

Here, $P(\mathbf{r}_1) = 3f \big/ 4\pi a^3$ is the probability of finding a particle at $\mathbf{r}_1$, $f$ is the particle volume fraction, and $\langle \mathbf{E}(\mathbf{r}) \,|\, \mathbf{r}_1 \rangle$ is the conditionally averaged electric field. Since the governing equations are linear and the medium is overall assumed to be macroscopically isotropic, $\langle g_p \mathbf{E} \rangle$ can be expressed as

$$\langle g_p \mathbf{E} \rangle(\mathbf{r}) = \Omega(f, k_{eff})\, f \langle \mathbf{E} \rangle(\mathbf{r}), \tag{5}$$

where $\Omega$ is a constant that depends on $k_{eff}$, $\phi$ and the microstructure. Combination of Eqs. 3 and 5 gives $\varepsilon_{eff}$ or equivalently $k_{eff}$ as follows:

$$k_{eff}^2 = k_m^2 - (k_p^2 - k_m^2)\Omega f. \tag{6}$$

The ensemble averaged Maxwell equation given by

$$(\nabla^2 + k_{eff}^2)\langle \mathbf{E} \rangle = \mathbf{0}, \tag{7}$$

is obtained by eliminating **H** from Eqs. 2a and 2b followed by a substitution of $k_{eff}^2$ from Eq. 6. $\Omega$ can be evaluated by solving Eq. 7 followed by substitution of $\langle \mathbf{E} \rangle$ and $\langle g_p \mathbf{E} \rangle$ into Eqs. 5 and 6. However, because $\Omega$ is a function of $k_{eff}$, Eqs. 6 and 7 need to be solved by an iterative numerical method for the evaluation of the zeros of the following function:

$$\Lambda(k_{eff}^2) = k_{eff}^2 - k_m^2 - f(k_p^2 - k_m^2)\Omega(k_{eff}^2) = 0. \tag{8}$$

## *A. The effective medium model*

To determine $\Omega$ we must determine the conditionally averaged **E** with one particle fixed and then evaluate the integral in Eq. 4. We shall use an effective medium model that has been shown to provide excellent predictions, consistent with rigorous computations that take into account multi-particle interactions of the conditionally averaged field and effective properties such as elastic moduli, attenuation and speed of acoustic waves, hydraulic permeability, effective viscosity and particle diffusivity in suspensions [33-38]. As illustrated in Fig. 1, in this model, the conditionally-averaged fields satisfy the governing equations for the suspending medium up to a distance $R$ from the center of the particle and the governing equations for the effective medium beyond that distance. $R$ is related to the static structure factor $S(\mathbf{0})$ as

$$\frac{R}{a} = \kappa = \left( \frac{1 - S(\mathbf{0})}{f} \right)^{\frac{1}{3}}, \tag{9}$$

where

$$S(\mathbf{0}) \equiv \int_0^\infty [P(\mathbf{r} \mid \mathbf{r}_1) - P(\mathbf{r}_1)] d\mathbf{r}. \tag{10}$$

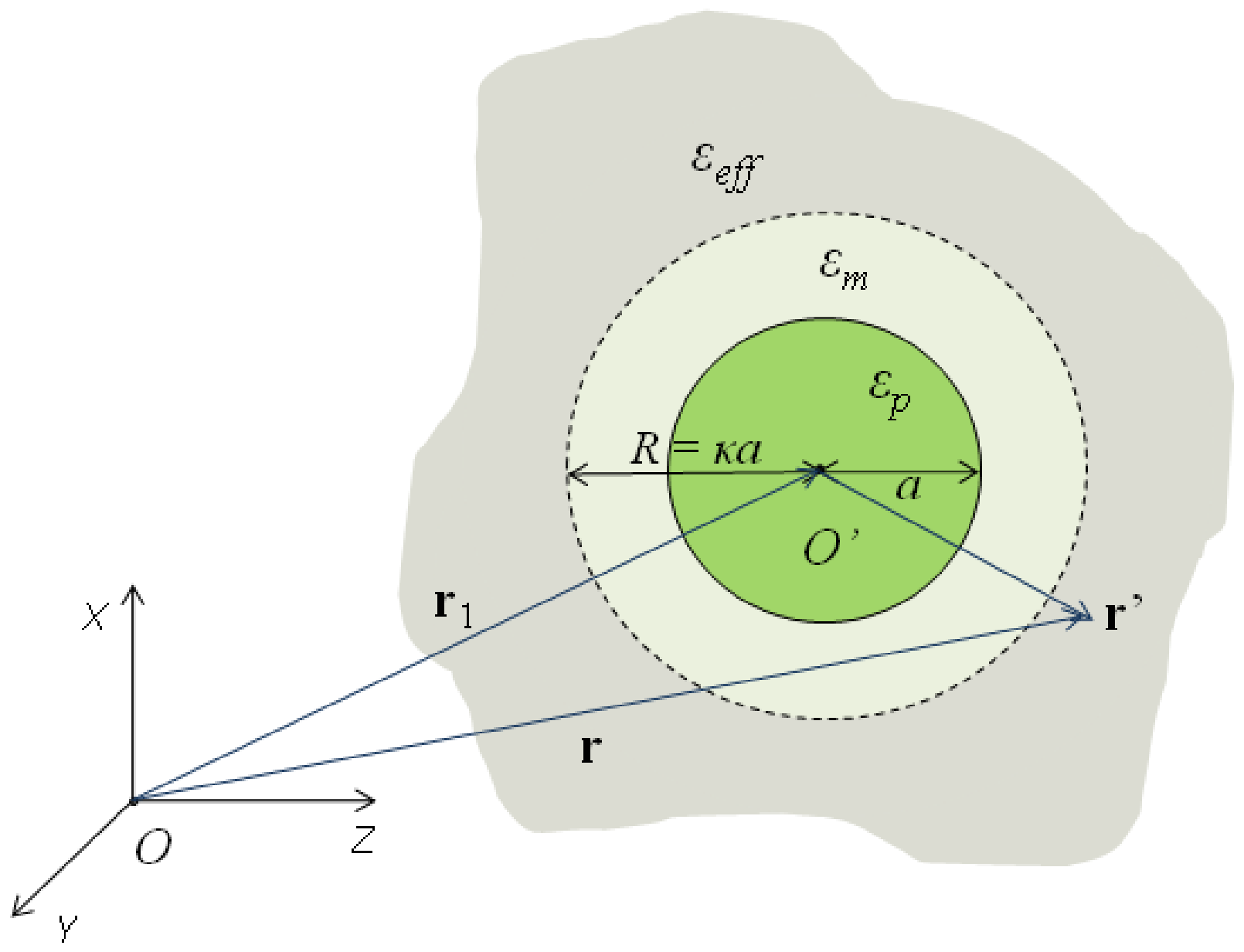

**Figure 1**: A schematic of the geometry considered for the EMT. The problem of finding the conditionally averaged field in a random medium was reduced to the problem of calculating the fields in this geometry. As $r \to \infty$, $\langle \mathbf{E} \rangle(\mathbf{r}) = \hat{\mathbf{x}} \exp(ik_{eff} z)$. The unconditionally averaged wave is assumed to be X-polarized in the present analysis. The choice of $\mathbf{r}_1$ is arbitrary for a given origin *O*.

In Eq. 10, $P(\mathbf{r} \mid \mathbf{r}_1)$ is the conditional probability density of the spheres. The quantity $S(\mathbf{0})$ can be interpreted as the integral of excess probability with respect to the uniform distribution. It needs to be accounted for if the random medium is to be replaced with a homogenous effective medium. In Fig. 1, *O* denotes the origin and *O'* denotes the centre of the particle and the particle is located at $\mathbf{r}_1$. Position vectors $\mathbf{r}'$ originate at *O'* such that $\mathbf{r} = \mathbf{r}_1 + \mathbf{r}'$.

As pointed out by Dodd et al. [35], the above choice of *R* is necessary to ensure that the conditionally averaged field has the correct behavior at large distances from the test particle for the problem of determining the averaged diffusivity of integral membrane proteins. Subsequent studies showed that the above choice also yields excellent estimates of the effective properties even when it has no rigorous basis (as e.g., see [33, 39]). Random suspensions with a hard-sphere potential have a non-zero $S(\mathbf{0})$ even in the dilute limit, which is accurately given by the Carnahan-Starling approximation [40] as:

$$S(\mathbf{0}) = \frac{(1-f)^4}{1+4f+4f^2-4f^3+f^4}. \quad (11)$$

As $f \rightarrow 0$, $\kappa = 2 - \frac{3}{2}f + O(f^2)$. On the other hand, well-separated dilute random arrays [41] have $P(\mathbf{r} \mid \mathbf{r}_1) = 0$ for $2 < r' < f^{-1/3}$, $S(\mathbf{0}) = 0$ and hence, $\kappa = f^{-1/3}$. We study both cases to elucidate the differences between them. Our effective medium theory is used to evaluate $\langle \mathbf{E}(\mathbf{r}) \mid \mathbf{r}_1 \rangle$ in Eq. 4.

## 3. Solution Technique

This section is devoted to a discussion on the determination of $\langle \mathbf{E}(\mathbf{r}) \mid \mathbf{r}_1 \rangle$ for $r' < a$, and subsequently, the numerical calculation of $\varepsilon_{eff}$. We first show that for composites with spheres that are small compared to the wavelength of the exciting light (or equivalently, as $ka \rightarrow 0$), $\langle \mathbf{E}(\mathbf{r}) \mid \mathbf{r}_1 \rangle$ and hence $\varepsilon_{eff}$ can be determined analytically by solving the electrostatic Maxwell equations (which is equivalent to setting $ka = 0$). Further, we show that both the electrostatic approximation and Maxwell wave equations can be reduced to Laplace equation for a scalar potential in the limit as $ka \rightarrow 0$. However, the boundary conditions for these two problems are different from each other. Since only a scalar potential is necessary to describe a static **E** field, the resulting EMT is referred to as the '*scalar EMT*'. Thereafter, an iterative method for the numerical evaluation of $\varepsilon_{eff}$ using Maxwell wave equations for arbitrary large values of *ka* is outlined. EMT based on Maxwell wave equations is referred to as the '*vector EMT*' as it involves solution of a vector Helmholtz equation. The boundary conditions for these two problems differ from each other as explained below.

### *A. Scalar EMT*

In this section, we will show that in the limit as $ka \rightarrow 0$, Maxwell wave equations given by $(\nabla^2 + k^2)\mathbf{E} = \mathbf{0}$ can be approximated by the electrostatic Maxwell equation given by $\nabla \times \mathbf{E}_S = 0$ for which $ka \equiv 0$ even though these equations require different boundary conditions. The electrostatic approximation is valid only for problems with a spherical symmetry and hence is not applicable to systems involving multiple spheres or non-spherical particles.

The static electric field $\mathbf{E}_S$ obeys $\nabla \times \mathbf{E}_S = 0$ and hence, can be represented as the gradient of a scalar potential $\Theta$ such that $\mathbf{E}_s = \nabla\Theta$. Therefore, $\nabla^2\Theta = 0$. Across an interface containing no free charge, $\Theta$ and the electric displacement $\mathbf{D}_s \equiv \varepsilon\mathbf{E}_s$ are continuous. Therefore, across a spherical interface between the particle and the medium shown in Fig. 1,

$$\Theta_p = \Theta_m \tag{12a}$$

and

$$\varepsilon_p \left.\frac{\partial \Theta}{\partial r}\right|_p = \varepsilon_m \left.\frac{\partial \Theta}{\partial r}\right|_m . \tag{12b}$$

At an interface with no free charge and current, **E** and **H** are required to have continuous tangential components. A comparative analysis can be performed by decomposing **E** and **H** into toroidal $\Psi$ and poloidal $\Phi$ scalar potentials in the following way [33, 42-43]. Let

$$\mathbf{E} = \nabla \times (\mathbf{r}\Psi) + \nabla \times \nabla \times (\mathbf{r}\Phi) \tag{13a}$$

and

$$\mathbf{H} = -i\omega\varepsilon \nabla \times (\mathbf{r}\Phi) + \frac{1}{i\omega\mu_0} \nabla \times \nabla \times (\mathbf{r}\Psi), \tag{13b}$$

where $\Psi$ and $\Phi$ are solutions to the scalar Helmholtz equation. Tangential components $E_\theta, E_\phi, H_\theta$ and $H_\phi$ can be expressed as follows [42-43]:

$$E_\theta = \frac{1}{\sin\theta}\frac{\partial \Psi}{\partial \phi} + \frac{1}{r}\frac{\partial^2 (r\Phi)}{\partial r\,\partial \theta}, \tag{14a}$$

$$E_\phi = -\frac{\partial \Psi}{\partial \theta} + \frac{1}{r}\frac{\partial^2 (r\Phi)}{\partial r\,\partial \phi}, \tag{14b}$$

$$H_\theta = \frac{-i\omega\varepsilon}{\sin\theta}\frac{\partial \Phi}{\partial \phi} + \frac{1}{i\omega\mu_0 r}\frac{\partial^2 (r\Psi)}{\partial r\,\partial \theta}, \tag{14c}$$

and

$$H_\phi = i\omega\varepsilon\frac{\partial \Phi}{\partial \theta} + \frac{1}{i\omega\mu_0 r}\frac{\partial^2 (r\Psi)}{\partial r\,\partial \phi}. \tag{14d}$$

Continuity of the above tangential components of **E** and **H** at the interface necessitate that

$$\Psi_p = \Psi_m, \tag{15a}$$

$$\left.\frac{\partial}{\partial r}(r\Psi)\right|_p = \left.\frac{\partial}{\partial r}(r\Psi)\right|_m, \tag{15b}$$

$$\varepsilon_p \Phi_p = \varepsilon_m \Phi_m \tag{15c}$$

and

$$\left.\frac{\partial}{\partial r}(r\Phi)\right|_p = \left.\frac{\partial}{\partial r}(r\Phi)\right|_m . \tag{15d}$$

As $ka \to 0$, $\nabla^2 \Psi = 0$ and $\nabla^2 \Phi = 0$. For a non-magnetic system, $\Psi$ is an indeterminable constant that does not contribute to **E** as seen from Eq. 13a. Hence, although the governing equations for the electrostatic and wave problems are identical, their boundary conditions differ from each other.

As discussed in §2A, our EMT is based on estimating the conditionally averaged fields using an effective medium model shown in Fig. 1. Hence, $\langle \Theta(\mathbf{r}) \,|\, \mathbf{r}_1 \rangle$ and $\langle \Phi(\mathbf{r}) \,|\, \mathbf{r}_1 \rangle$ are obtained by the solution of Laplace equations for $\Theta$ and $\Phi$ subject to the boundary conditions given in Eqs. 13 and 15 respectively. In order to solve the Laplace problem, the unconditionally averaged far-field given by $\langle \mathbf{E} \rangle = \hat{\mathbf{x}} \exp(i\mathbf{k}_{eff} \cdot \mathbf{r})$ can be expressed in terms of the first Laplace harmonic in the following way. One may consider rotating the coordinate system in Fig. 1 about the y-axis such that $\hat{\mathbf{x}}$ is replaced by $\hat{\mathbf{z}}$ and hence **E** points in the direction of the zenith. Since, $\mathbf{r} = \mathbf{r}_1 + \mathbf{r}'$, $\exp(i\mathbf{k}_{eff} \cdot \mathbf{r}) \to \exp(i\mathbf{k}_{eff} \cdot \mathbf{r}_1)$ as $r' \to 0$. Hence, far field scalar potentials $\langle \Theta \rangle$ and $\langle \Phi \rangle = \exp(i\mathbf{k}_{eff} \cdot \mathbf{r}_1) r' \cos\theta$. $\langle \mathbf{E}(\mathbf{r}) \,|\, \mathbf{r}_1 \rangle$ and $\langle \mathbf{E}_S(\mathbf{r}) \,|\, \mathbf{r}_1 \rangle$ are given by $\nabla \langle \Theta(\mathbf{r}) \,|\, \mathbf{r}_1 \rangle$ and $\nabla \langle \Phi(\mathbf{r}) \,|\, \mathbf{r}_1 \rangle$ respectively. We find that the coefficient of the first regular harmonic is identical for $\Theta$ and $\Phi$ and consequently $\langle \mathbf{E}(\mathbf{r}) \,|\, \mathbf{r}_1 \rangle = \langle \mathbf{E}_S(\mathbf{r}) \,|\, \mathbf{r}_1 \rangle$

$$= \exp(i\,\mathbf{k}_{eff} \cdot \mathbf{r}_1) \frac{3\varepsilon_m}{\varepsilon_p + 2\varepsilon_m} \hat{\mathbf{z}},\ r' < a \tag{16}$$

for a sphere in an infinite matrix. A similar procedure can be employed to show that

$$\langle \mathbf{E}_S(\mathbf{r}) \,|\, \mathbf{r}_1 \rangle = \exp(i\,\mathbf{k}_{eff} \cdot \mathbf{r}_1) \frac{9\varepsilon_m \varepsilon_{eff}}{(\varepsilon_p + 2\varepsilon_m)(2\varepsilon_{eff} + \varepsilon_m) - 2\kappa^{-3}(\varepsilon_p - \varepsilon_m)(\varepsilon_{eff} - \varepsilon_m)} \hat{\mathbf{z}},\ r' < a \tag{17}$$

for a sphere an *effective medium* shown in Fig. 1. The exponential term in Eqs. 16 and 17 is a phase factor that depends on the location of the sphere. Eqs 4, 5 and 6 can be used in that order to obtain an expression for $\varepsilon_{eff}$. Required algebraic manipulations are discussed in §3B below. Exact expressions for $\varepsilon_{eff}$ are presented §4A.

## *B. Vector EMT*

The determination of $\langle \mathbf{E}(\mathbf{r}) \,|\, \mathbf{r}_1 \rangle$ for a finitely large value of *ka* requires the solution of the vector Helmholtz equation for a 2-layer sphere geometry shown in Fig. 1. **E** inside a particle can be found by utilizing a multipole expansion. The solution given by Hightower and Richardson [44] was adapted here to obtain the following relations for $\langle \mathbf{E}(\mathbf{r}) \,|\, \mathbf{r}_1 \rangle$ for $r' < a$:

$$\langle E_r(\mathbf{r}) \,|\, \mathbf{r}_1 \rangle = \exp(i\mathbf{k}_{eff} \cdot \mathbf{r}_1) \frac{-i \sin\theta' \cos\phi'}{k_p^2 r'^2} \sum_{n=1}^{\infty} d_n i^n (2n+1) \pi_n(\theta') \psi_n(k_p r'), \qquad (18a)$$

$$\begin{aligned} \langle E_\theta(\mathbf{r}) \,|\, \mathbf{r}_1 \rangle &= \exp(i\mathbf{k}_{eff} \cdot \mathbf{r}_1) \frac{\cos\phi'}{k_p r'} \\ &\times \sum_{n=1}^{\infty} i^n \frac{(2n+1)}{n(n+1)} [c_n \pi_n(\theta') \psi_n(k_p r') - i d_n \tau_n(\theta') \psi_n'(k_p r')] \end{aligned} \qquad (18b)$$

and

$$\begin{aligned} \langle E_\phi(\mathbf{r}) \,|\, \mathbf{r}_1 \rangle &= \exp(i\mathbf{k}_{eff} \cdot \mathbf{r}_1) \frac{\sin\phi'}{k_p r'} \\ &\times \sum_{n=1}^{\infty} - i^n \frac{(2n+1)}{n(n+1)} [c_n \tau_n(\theta') \psi_n(k_p r') - i d_n \pi_n(\theta') \psi_n'(k_p r')]. \end{aligned} \qquad (18c)$$

In Eqs. 18a-c, primed coordinates $(r', \theta', \phi')$ are with respect to the origin *O'*, *n* is the order of the multipole, $\pi_n = \dfrac{P_n^1(\cos\theta)}{\sin\theta}$ and $\tau_n = \dfrac{dP_n^1(\cos\theta)}{d\theta}$ are the polar angle dependent functions related to the associated Legendre polynomials $P_n^1$ of degree one, $\psi_n(z) \equiv z j_n(z)$ are Riccati-Bessel functions associated with the spherical Bessel functions $j_n(z)$ and $c_n$ and $d_n$ are the corresponding Mie coefficients [21]. The expressions of $c_n$ and $d_n$ are given in Appendix A. In general, both $c_n$ and $d_n$ are functions of *a* and $k_{eff}$ [21].

Bessel functions and their derivatives were calculated using established iterative techniques [45]. The volume integral of $\langle \mathbf{E}(\mathbf{r}) \,|\, \mathbf{r}_1 \rangle$ over the particle volume was determined to evaluate $\langle g_p \mathbf{E} \rangle$ and subsequently Ω using Eqs. 4 and 5. Phase factors $\exp(i\mathbf{k}_{eff} \cdot \mathbf{r}_1)$ in Eqs. 18 were expressed as $\exp(i\mathbf{k}_{eff} \cdot \mathbf{r})\exp(-i\mathbf{k}_{eff} \cdot \mathbf{r}')$ so that $\exp(i\mathbf{k}_{eff} \cdot \mathbf{r})$ in the right-hand side of Eq. 5 would cancel out with the left-hand side. A two dimensional composite Simpson's rule [46] was used since the analytical evaluation of the integral in Eq. 4 over $r$ and $\theta$ was not possible due to the presence of the exponential term $\exp(-i\mathbf{k}_{eff}\cdot\mathbf{r}')$. Only the $x$-component of $\langle \gamma_p \mathbf{E}\rangle$ was found to be non-zero, consistent with the isotropic nature of the effective medium. $k_{eff}$ and equivalently $\varepsilon_{eff}$ were calculated by finding the zeros of $\Lambda$ in Eq. 7 using Newton-Raphson iterations. Since, $\Lambda$ was not necessarily analytic in the complex variable $k_{eff}$, it was treated as a function of two variables which were the real and imaginary parts of $k_{eff}$ necessitating the use of a two dimensional Newton-Raphson method [47]. For large $f$, Eq. 8 permitted multiple solutions that were close to one other. Hence, the solution corresponding to the limit as $f \to 0$ was traced by using zero order continuation. The procedure was repeated for $\lambda$ values in the visible range (300-800 nm). Permittivity data for noble metals was obtained from Ref. [48]. All computations were performed using MATLAB.

## 4. Results and discussion

### A. Scalar EMT

In the limit as $f \to 0$, $\varepsilon_{eff} \to \varepsilon_m$. Hence, due to the absence of particle-effective medium coupling, $\langle \mathbf{E}(\mathbf{r}) \,|\, \mathbf{r}_1 \rangle$ is given by Eq. 16. The corresponding $\varepsilon_{eff}$ is given by the following equation:

$$\varepsilon_{eff}/\varepsilon_m = 1 + 3\beta f, \quad \beta = \left(\frac{\varepsilon_p - \varepsilon_m}{\varepsilon_p + 2\varepsilon_m}\right). \tag{19}$$

Here, $\beta$ is the electric polarizability per unit volume for a small sphere. As will be shown later in this section, the linear dependence shown in Eq. 19 is a general result that is independent of the mircostructure since $\beta$ is a material property. In the case of a finitely large $f$ and arbitrary $\kappa$, $\langle \mathbf{E}(\mathbf{r}) \,|\, \mathbf{r}_1 \rangle$ is given by Eq. 17 for which $\varepsilon_{eff}$ is found to be the following:

$$\varepsilon_{eff}/\varepsilon_m = 1 + \frac{9\varepsilon_{eff}}{(2\varepsilon_{eff} + \varepsilon_m) - 2\kappa^{-3}\beta(\varepsilon_{eff} - \varepsilon_m)}\gamma, \gamma \equiv \beta f. \tag{20}$$

For a well-separated system, $\kappa = f^{-1/3}$, which in conjunction with Eq. 20 gives the following result:

$$\varepsilon_{eff} / \varepsilon_m = \frac{1+2\gamma}{1-\gamma} = 1+3\gamma+3\gamma^2+O(\gamma^3). \tag{21}$$

Eq. 21 is identical to the classical Maxwell-Garnett theory (MGT) which is also a scalar EMT in which the presence of other particles is accounted through the modification of the averaged far-field [28, 49]. The lower bound, $\kappa = 1$, can be substituted in Eq. 20 to give the well-known Bruggeman mixing rule (BMR):

$$f\left(\frac{\varepsilon_p - \varepsilon_{eff}}{\varepsilon_p + 2\varepsilon_{eff}}\right) + (1-f)\left(\frac{\varepsilon_m - \varepsilon_{eff}}{\varepsilon_m + 2\varepsilon_{eff}}\right) = 0, \tag{22}$$

that is based on a symmetric mixing approach for the inclusions and matrix phase. As a consequence, BMR can model percolation effects [27]. A similar concentric-shell model with a variable shell thickness was also proposed by Hashin and Shtrikman [27]. However, the dependence of the shell thickness parameter $\kappa$ on the microstructure was not demonstrated. MGT and BMR can be seen as the upper and lower Hashin-Strikman bounds of the scalar EMT. Garcia et al., among others, have derived self-consistent mixing rules for ternary plasmonic composites based on Hashin-Strikman formalism [50]. Within the framework of the EMT presented in this work, $\kappa$ is a *physical parameter* that can be determined from the structure factor (or equivalently the radial distribution function) of the composite. Conversely, if $\kappa$ were to be determined by fitting spectroscopic data to the EMT predictions, it can be used to glean microstructure information regarding the distribution of particles within the composite.

For a random system, $\kappa$ can be expanded about $f = 0$ using Eqs. 9 and 11 to give $\kappa = 2 - \frac{4}{3}f + O(f^2)$. This can be substituted in Eq. 20 to obtain the following expansion for $\varepsilon_{eff}$ valid for $\gamma << 1$:

$$\varepsilon_{eff} / \varepsilon_m = 1+3\gamma+\frac{3}{4}(\beta+4)\gamma^2+O(\gamma^3). \tag{23}$$

Note that all scalar EMTs based on Eq. 20 indeed yield $\varepsilon_{eff} = \varepsilon_p$ for $f = 1$.

## *I. Resonance conditions*

A particle undergoes an electric resonance when $\langle \mathbf{E}(\mathbf{r}) | \mathbf{r}_1 \rangle$ is singular. For a finitely large $f$, however, a different resonance condition that takes into account the particle-effective medium coupling effect will result. Hence, in this section, a discussion on the conditions under which scalar $\varepsilon_{eff}$ shows a resonance is presented.

Resonance of a single sphere requires that

$$\left|\beta\right| \to \infty \text{, or equivalently, } \varepsilon_p = -2\varepsilon_m \tag{24}$$

as can seen from Eqs. 16 and 19. Presence of an effective medium, on the other hand, leads to a condition that the denominator in Eqs. 17 and 20 vanish. Hence, one obtains a resonance condition given by:

$$2\beta\left(\frac{\varepsilon_{eff} - 1}{2\varepsilon_{eff} + 1}\right) = \kappa^3 . \tag{25}$$

In the specific case of a well-separated system (or MGT), substitution of $\kappa = f^{-1/3}$ in Eq. 25 gives the following resonance condition:

$$\beta \approx \frac{1}{f}. \tag{26}$$

The above equation can also be obtained by letting the denominator in Eq. 21 to be zero. Similar to the way by which Eq. 26 was obtained, a substitution of $\kappa = 2 - \frac{4}{3}f + O(f^2)$ in Eq. 25 under the limit as $\gamma \to 0$, leads to the following resonance condition for random hard-sphere composites:

$$\beta \approx \frac{2}{\sqrt{f}}. \tag{27}$$

As seen from Eqs. 26 and 27, for dense systems the resonance wavelength is different from that of a single particle. Hence, the scaling of $\beta$ with $f$ depends on the microstructure of the system through the static structure factor. This trend is consistent with the reported red-shift in plasmon resonance for ion-implanted composites [13]. We note that substitution of $\kappa = 1$ in Eq. 25 in the limit as $\gamma \to 0$ gives the resonance condition in the 'Bruggeman limit' as $\beta \approx 1/(2\sqrt{f})$.

Since the equations for dense systems (Eqs. 17, 20 and 24) include particle-effective medium coupling, they are implicit in $\varepsilon_{eff}$. Conversely, the inverse problem, i.e. when $\varepsilon_{eff}$ is given and $\varepsilon_p$ or $f$ are unknown, is explicit in all cases. Hereafter, we only consider the solution of Eq. 20 that approaches unity as $f \rightarrow 0$.

## *II. Ag Plasmonic composite*

Resonant plasmonic nanospheres made of noble metals such as Ag, Au and Cu have an $\varepsilon_p$ that has a large negative real part and a small positive imaginary part for the visible range of the electromagnetic spectrum as shown in Fig. 2. Their $\beta$ values show a resonance as a consequence. We consider Ag in our further discussion since it has the smallest $\Im(\varepsilon_p)$ over a broad wavelength range and hence the most prominent $\beta$. Further, a medium with a relatively large $\varepsilon_m$ can shift this resonance to the red region and make it more prominent. For this work we will consider $\varepsilon_m = 7$. Semiconductors such as ZnO, Si, $TiO_2$ etc. have similar values of $\varepsilon_m$. Values of $\beta$ for Ag spheres in an $\varepsilon_m = 7$ medium are shown in Fig. 3.

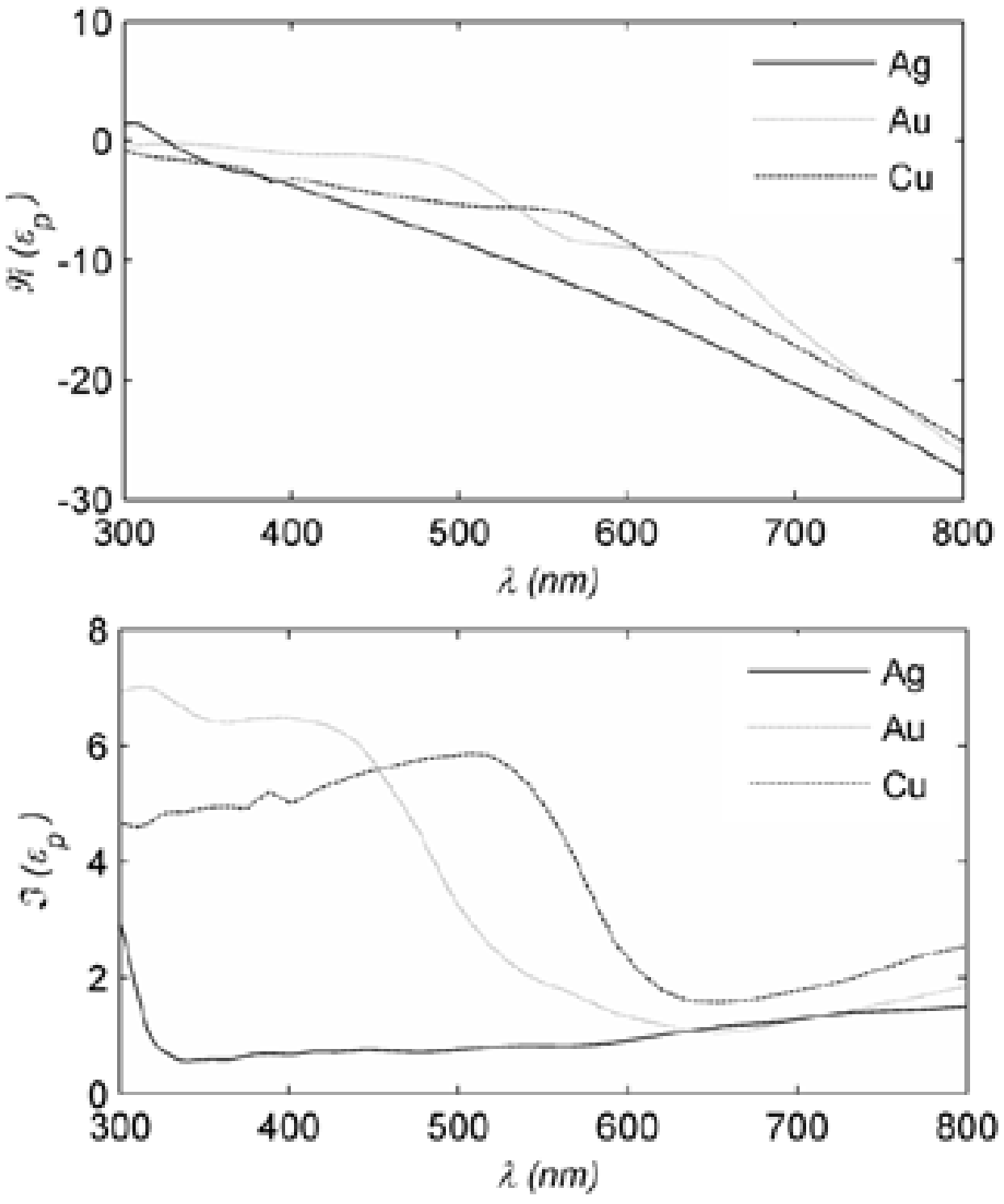

**Figure 2**: Real and imaginary parts of the permittivity of high optical conductivity metals Ag, Au and Cu that are considered in this work. Permittivity data is taken from Ref. [48]. Ag has the lowest imaginary permittivity over a broad range of wavelengths.

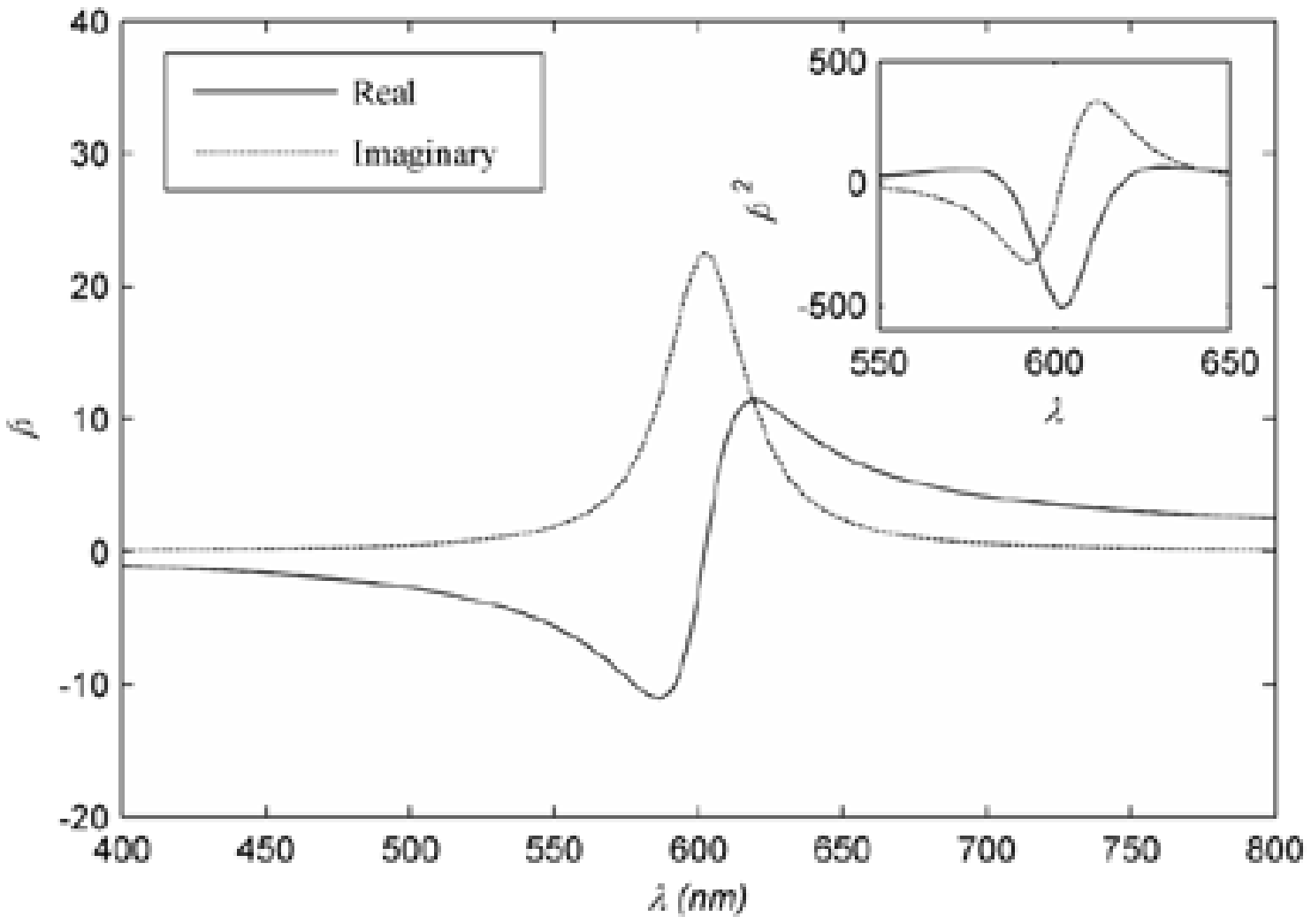

**Figure 3**: $\beta$ for Ag spheres in a $\varepsilon_m = 7$ medium. Resonance occurs for $\lambda \approx 600$ nm and $\beta \approx 23i$. $\Re(\beta)$ is a small number away from resonance and $\Im(\beta)$ changes sign from negative to positive on moving from blue to red regions about resonance.

The permittivity of high conductivity metals such as Ag can be evaluated approximately by using the Drude model given by $\varepsilon_p = 1 - \dfrac{\omega_p^2}{\omega^2 + i\omega\omega_c}$ [51]. Here, plasma frequency $\omega_p \approx 2.321 \times 10^{15}$ Hz and collision frequency $\omega_c \approx 5.513 \times 10^{12}$ Hz for Ag in the visible range [52]. The peak in $\Re(\beta)$ occurs at the resonance frequency given by $\omega_R = \dfrac{\omega_p}{\sqrt{2\varepsilon_m + 1}}$ or equivalently resonance wavelength $\lambda_R = \dfrac{2\pi c\sqrt{2\varepsilon_m + 1}}{\omega_p}$ where *c* is the speed of light in vacuum. Hence, the resonance wavelength for small spheres scales as $\sqrt{2\varepsilon_m + 1}$ resulting in a red-shift as $\varepsilon_m$ is increased. The peak value $\beta_R$ at resonance is given by $\beta_R = -\left(\dfrac{\varepsilon_m - 1}{2\varepsilon_m + 1}\right) + i\dfrac{\omega_p}{\omega_c}\dfrac{3\varepsilon_m}{(2\varepsilon_m + 1)^{3/2}}$ for a Drude metal.

The real and imaginary parts of $\beta$ represent the reactive and dissipative components respectively. At resonance ($\lambda \approx 600$ nm), $\Re(\beta)$ is a large positive number and $\Im(\beta) = 0$ as shown in Fig. 3. As $|\gamma| \to 0$, the dissipative term for all scalar EMTs is given by $\Re(\varepsilon_{eff}/\varepsilon_m) \approx 3f\,\Re(\beta)$. The quadratic term for random systems is $-\dfrac{3}{4}[\Re(\beta)]^3 f^2$. As a result, $\Im(\varepsilon_{eff})$ is highly non-linear in $\phi$. For $\beta$ shown in Fig. 3, $\Im(\varepsilon_{eff}/\varepsilon_m) \approx 69\phi - 9125.25\phi^2$ at resonance. In contrast, it is identically zero for a well-separated composite. Hence, quadratic coupling does not lead to any loss in a well-separated composite. $\Re(\varepsilon_{eff}/\varepsilon_m)$ is identically one under resonance for a random system. Hence, the reaction originates only from the medium. For a well-separated system, quadratic and even powered coupling contributes to the reaction such that $\Re(\varepsilon_{eff}/\varepsilon_m) \approx 1 - 3[\Im(\beta)\phi]^2$.

At off-resonance, $\Re(\beta)$ is greater than zero for $\lambda > \lambda_R$ and less than zero for $\lambda < \lambda_R$. $\Im(\beta)$ is smaller in comparison for the most part. For instance, at $\lambda \approx 680$ nm, $\beta = 4.973 + 0.973i$; and at $\lambda \approx 520$ nm, $\beta = -3.546 + 0.816i$. The red-regions exhibit a "concentration resonance" which occurs when the resonance condition given in Eq. 24 is satisfied. In case of a well-separated system, the condition is $\beta\phi = 1$ and for a random system, it is $\beta^2\phi \approx 4$ as given in Eqs. 25 and 26. Fig. 4 shows $\Im(\varepsilon_{eff}/\varepsilon_m)$ as a function of $\phi$ for $\lambda \approx 680$ nm. A peak appears at $\phi \approx 13\%$ for a random system and at $\phi \approx 20\%$ for a well-separated system. The peak for a random system is less prominent compared to that for a well-separated system. The resonance concentration shifts to higher values of $\phi$ as $\varepsilon_m$ is decreased.

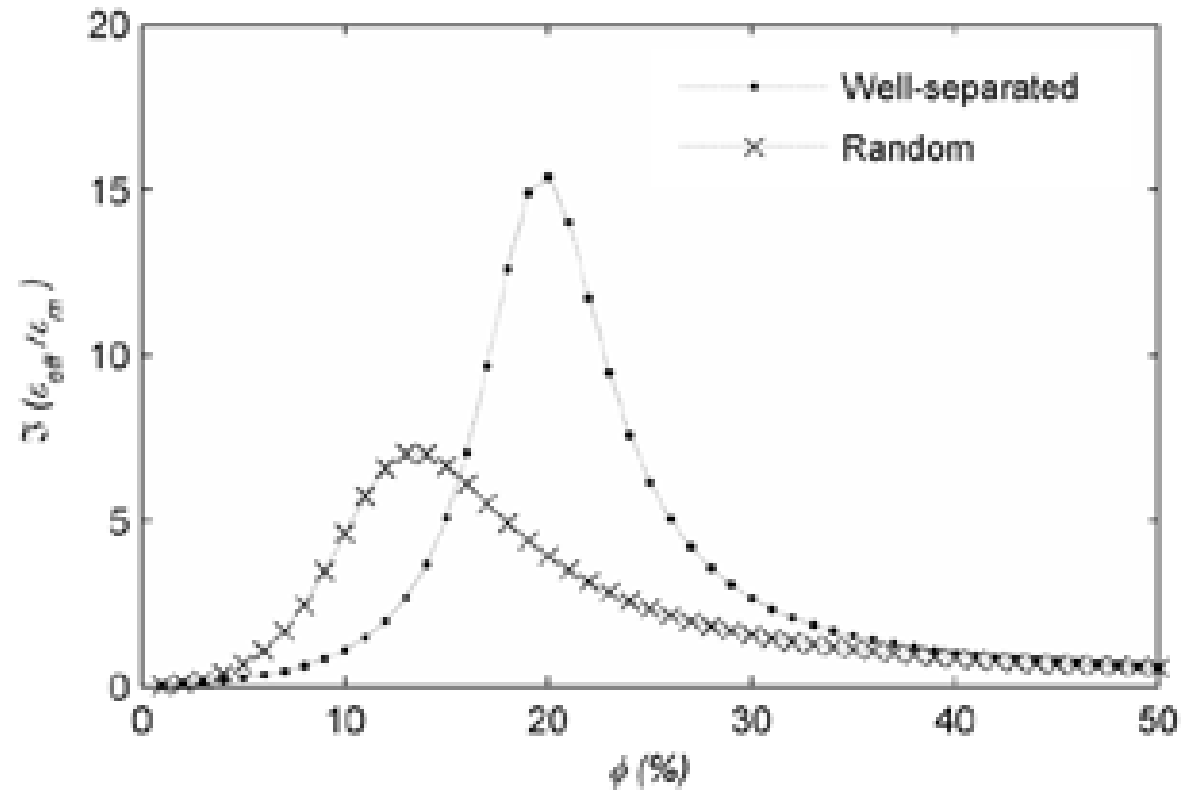

Figure 4: $\Im(\varepsilon_{eff})$ predicted by the scalar EMT for random and well-separated microstructures. Here, $\lambda \approx 680$ nm and $\beta = 4.973 + 0.973i$. $\beta$ has a resonance peak at $\lambda \approx 600$ nm as shown in Fig. 3.

Fig. 5 shows the effect of $\phi$ on $\varepsilon_{eff}$. It is evident that a well-separated system couples more intensely with the incident field as compared to a random one. Peaks corresponding to random systems occur for $\lambda$ values that are larger than those for well-separated systems due to the difference in the scaling of $\beta$ with respect to $\phi$ under resonance as shown in Eqs. 26 and

27. The quadratic coefficient for a random system is $\frac{3}{4}\beta^3 + 3\beta^2$ (Eq. 22) in comparison to $3\beta^2$ (Eq. 21) for a well-separated system. The coefficients of the cubic and higher order terms in $f$ also depend on higher powers of $\beta$ for a random system. Hence, the resonance condition for the coefficient of the quadratic term in the power series expansion of $\varepsilon_{eff}(f)$ will depend on that of the linear term while the resonance condition for cubic coefficient will depend on those of the quadratic and linear terms, etc. As a consequence, the peak in $\Re(\varepsilon_{eff})$ for random systems is not symmetric about its maximum and is broader in comparison to that for a well-separated system or a single sphere.

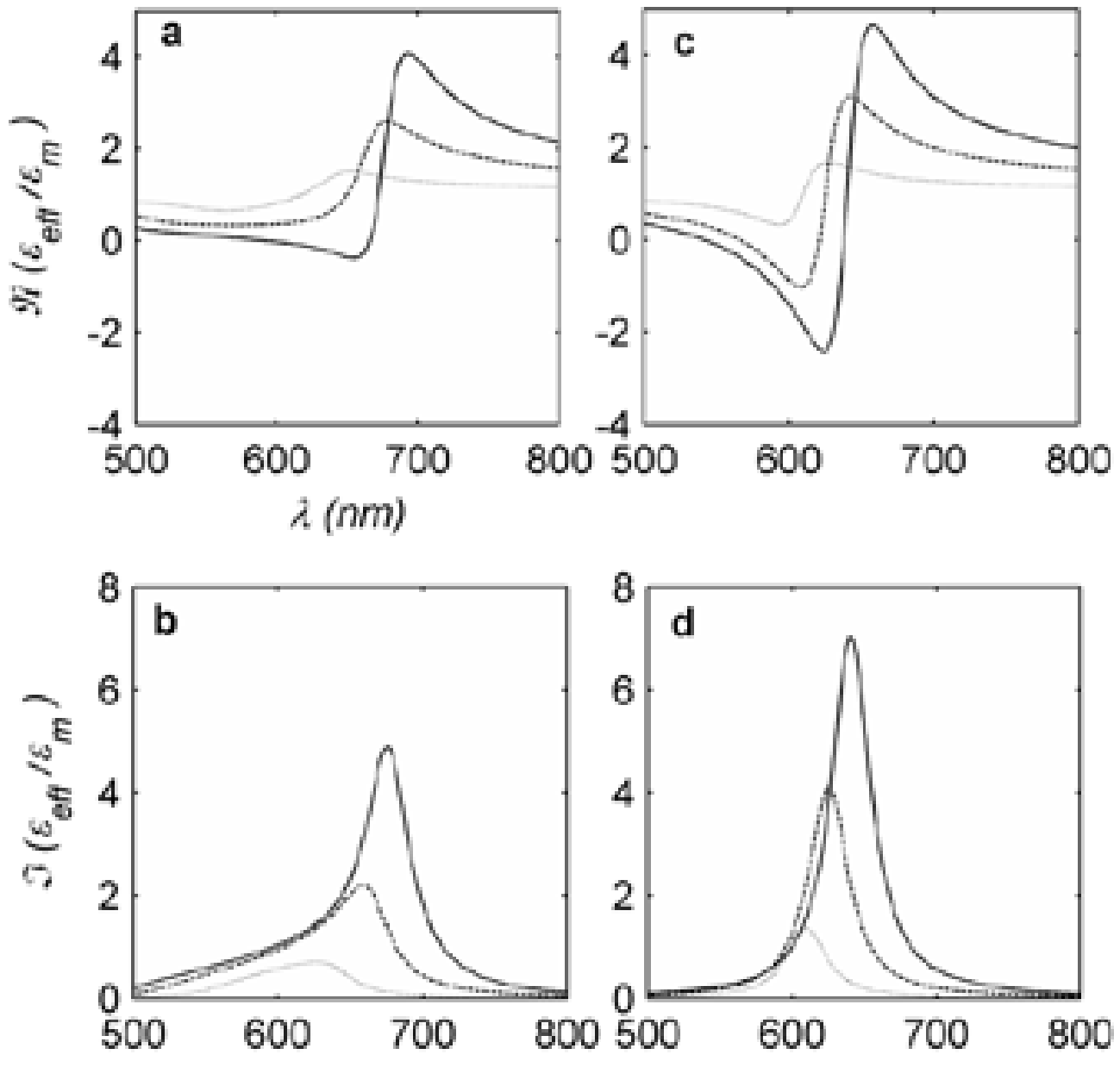

Figure 5: $\varepsilon_{eff}$ for a composite with Ag NPs in an $\varepsilon_m = 7$ medium calculated with the scalar EMT. Random (a, b) and well-separated random composites (c, d) for $f = 2\%$ (solid), 6% (dashed) and 10% (dotted) were considered. The resonance peak is more red-shifted and broad for a random system. A well separated system shows a more intense resonance with a symmetric peak in comparison. Stronger coupling in a random system leads to a tail in the blue region.

## *B. Vector EMT*

This section describes the effect of *ka* on numerically calculated values of $\varepsilon_{eff}$. An analysis of vector EMT in the limit as $f \to 0$ and $ka \to 0$ is first presented. Resonance analysis in the limit as $\phi \to 0$ is presented in §4B-I followed by a discussion on the numerical results for finitely large $\phi$ and $ka$ will follow in §4B-II.

In the vector EMT, $\langle \mathbf{E}(\mathbf{r}) \,|\, \mathbf{r}_1 \rangle$ is not a constant vector but depends on the toroidal (corresponding to coefficients $c_n$) and poloidal (corresponding to coefficients $d_n$) multipoles shown in Eqs. 18. Only poloidal modes are capable of generating $\langle E_r \,|\, \mathbf{r}_1 \rangle$ which only toroidal modes generate $\langle H_r \,|\, \mathbf{r}_1 \rangle$. Hence, we will refer to the poloidal modes as the electric modes and the toroidal modes as magnetic modes following the standard convention [43].

For subsequent analysis, the following expansion is used to express the dependence of $\varepsilon_{eff}$ on $\phi$:

$$\varepsilon_{eff} / \varepsilon_m = 1 + A\phi + B\phi^2 + O(\phi^3). \tag{28}$$

As in §3A, only the linear and quadratic coefficients are of interest here. Coefficients $A$ and $B$ in Eq. 28 depend on $ka$, $\varepsilon_p$, and $\varepsilon_m$. Riccati-Bessel functions in Eqs. 18 can be expanded into a Taylor series as $\psi_n(z) = C_n z^{n+1} + O(z^{n+3})$ where $C_n = \dfrac{\sqrt{\pi}}{4\Gamma(n+3/2)}$, where $\Gamma$ denotes the gamma function. Hence, as $z \to 0$ or equivalently $k_p a \to 0$, the terms corresponding the $c_1$ mode in Eqs. 18 are $O(z)$, while those corresponding to the $d_1$ mode are $O(1)$. As a result, only the $d_1$ term contributes significantly as $z \to 0$. In the limit as $\phi \to 0$, only the linear coefficient $A$ in Eq. 28 is relevant irrespective of the microstructure. $\langle \mathbf{E}(\mathbf{r}) \,|\, \mathbf{r}_1 \rangle$ from Eqs. 18 can be used together with Eq. 4 and 5 to obtain the parameter $\Omega$. It can be shown that $\Omega = d_1$ and as $k_p a \to 0$, where $d_1$ is given by $d_1 = \dfrac{3}{(\varepsilon_p + 2)} \left[ 1 + i\dfrac{2}{3}\beta (k^* a)^3 \right] + O\left[(k^* a)^6\right]$. Subsequently, $A$ can obtained by using Eq. 6 as:

$$A = 3\beta + i2\beta^2 (k^* a)^3 + O\left[(k^* a)^6\right], \tag{29}$$

where, $k^* = k_p / n_{eff}$. Note that as $\phi \to 0$, $n_{eff} \to n_m$ and hence, $k^*$ can be modified appropriately in the dilute limit. Mallet et al. [29] have recently re-derived MGT for finitely large particles that can exhibit scattering using rigorous Foldy-Lax multiple scattering equations (see Eq. 22 in [29]). The linear term is identical to the one obtained here in Eq. 29. The cubic dependence on the particle radius implies that the size effect becomes significant only when $\left|\beta^2\left(k^*a\right)^3\right|$ is $O(1)$. Since $\beta^2$ is a large negative number at resonance as shown in Fig. 3 inset, size effects become significant even for relatively small radii for resonant systems. For example $\left|\beta^2\right| = 500$ at resonance for an Ag sphere shown in Fig. 3. Hence, $\left|\beta^2\left(k^*a\right)^3\right| = 1/2$ for $\left|k^*a\right| \approx 0.063$. Here, since $k^* = 0.0005 + 0.0147i$ nm$^{-1}$, a relatively small radius of $a \approx 34$ nm can significantly affect $\varepsilon_{eff}$ even in the dilute limit.

## *I. Resonance conditions*

Resonance occurs when $\left\langle \mathbf{E}(\mathbf{r}) \,|\, \mathbf{r}_1 \right\rangle$ given by Eqs. 18 is singular. In turn, this requires that the coefficients $c_n$ and $d_n$ are singular. Magnetic resonances represented by singular $c_n$s do not occur in plasmonic systems since $\mu_p = \mu_m = \mu_0$. Vector EMT exhibits only electric multipole resonances that correspond to $d_n$ that are given in Appendix A. The conditions are complicated for Helmholtz multipoles as they involve the Riccati-Bessel functions. However, the underlying physical aspects can be appreciated by utilizing the simpler Laplace multipoles [53]. The $n^{th}$ Laplace multipole has a size independent polarizability $\beta_n$ defined as:

$$\beta_n \equiv \frac{\varepsilon_p - \varepsilon_m}{\varepsilon_p + \dfrac{n+1}{n}\varepsilon_m}. \tag{30}$$

A dipole resonance requires that $\beta_1 \to \infty$ or equivalently $\varepsilon_p = -2\varepsilon_m$; quadrupole resonance occurs when $\beta_2 \to \infty$ or $\varepsilon_p = -\frac{3}{2}\varepsilon_m$ and so on [53]. Hence, higher order multipoles become resonant at smaller negative values of $\varepsilon_p$, or equivalently for smaller values of $\lambda$ as inferred from the $\varepsilon_p - \lambda$ curve shown in Fig. 2. Mie coefficients $A_n$, $a_n$ and $d_n$ are polynomials of $k^*a$ and $\beta_n$ in the limit as $k^*a \to 0$ such that higher order $\beta_n$ become significant only for an $O(1)$ $k^*a$ [21, 54]. For an arbitrary $k^*a$, Helmholtz multipole electric polarizabilities $d_n$ depend on $k^*a$ [54] such that their resonance peak red-shifts with $k^*a$. Hence, both $A$ and $B$ red-shift as $k^*a$ increases.

The above mentioned physical trends in the resonance conditions are also seen in the numerical results obtained for the vector EMT. Fig. 6 shows $A$ as a function of $\lambda$ for diameters $d_p$ = 10, 30, 50 and 100 nm. Here, $k^*a \approx \frac{0.45 d_p \sqrt{\varepsilon_p}}{\lambda}$, where $d_p$ and $\lambda$ are in nm. $A$ was calculated by fitting the data for $\varepsilon_{eff}$ obtained for $\phi = 0.01\%$ to Eq. 28 as:

$$A = \frac{1}{\phi}\left(\frac{\varepsilon_{eff}}{\varepsilon_m} - 1\right). \tag{31}$$

For a small value of $d_p$ such as 10 nm, only the dipole mode is significant. Hence, $A = 3\beta$. Relatively larger particles, e.g. $d_p$ = 50 nm, show a significant quadrupole resonance. The dotted curves in Fig. 6 shows a quadrupole peak at $\lambda \approx 580$ nm. A further larger $d_p$ = 100 nm results in an octupole peak as depicted in the insets of Fig. 6. Each peak red-shifts for larger $d_p$. For example, the quadrupole peak for $d_p$ = 100 nm occurs at $\lambda \approx 650$ nm. **E** becomes highly localized at the particle surface for relatively large $k^*a$ values. Consequently, the magnitude of $A$ and $\varepsilon_{eff}$ are diminished for relatively large $d_p$ values as shown in Fig. 6. In the large size limit ($k^*a \rightarrow \infty$), $A$ = 0. Further an inspection of Eqs. 3 and 18 shows that $\varepsilon_{eff} = \varepsilon_m$ for arbitrarily large $\phi$, in the limit as $k^*a \rightarrow \infty$. This is consistent with the ray optics scenario.

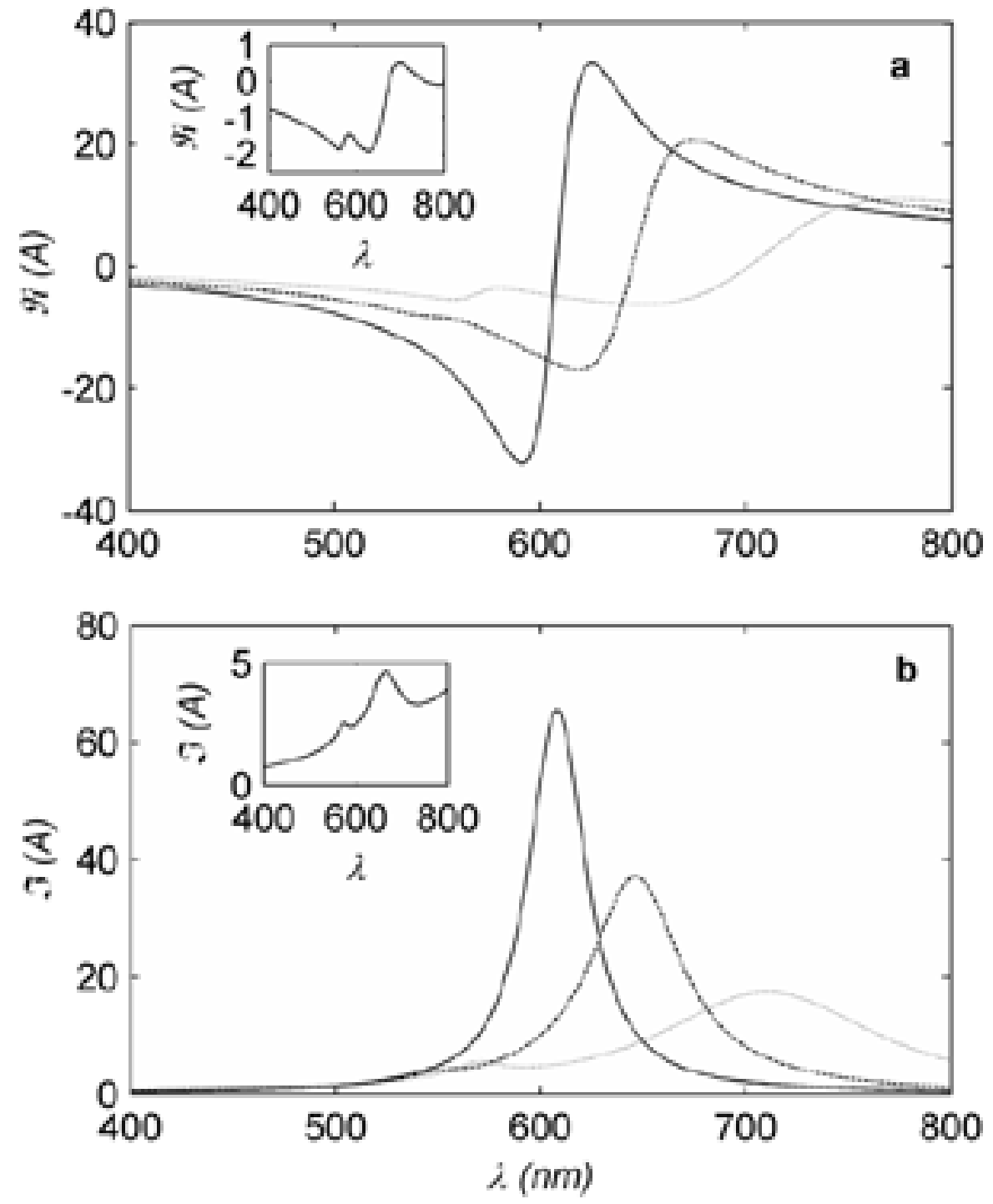

**Figure 6**: Linear coefficient *A* for composites in an $\varepsilon_m = 7$ matrix with Ag NPs with diameters $d_p = 10$ (solid), 30 (dashed), 50 (dotted) and 100 nm (inset). Here, $k^* a \approx \frac{0.45 d_p \sqrt{\varepsilon_p}}{\lambda} < 1$ only for the blue curve. Quadrupolar and octupolar resonance peaks are present for large particles as seen in the curves in the insets. Dipole resonance is most prominent and red-shifts as $d_p$ is increased. The linear coefficient becomes less significant for large particles as they screen most of the **E** field from their interior.

By neglecting cubic and higher order terms in Eq. 28, the coefficient *B* was obtained using the following expression with $f = 1\%$:

$$B = \frac{1}{f^2}\left(\frac{\varepsilon_{eff}}{\varepsilon_m} - 1 - Af\right). \tag{32}$$

Fig. 7 shows the values of *B* calculated for random ($\kappa$ given by Eqs. 9 and 11) and well-separated ($\kappa = f^{-1/3}$) random systems. *B* represents the strength of interparticle coupling. Hence, random systems have a larger *B*. The $n^{th}$ multipole decay as $r^{-(n+2)}$ in general. Hence, dipoles can couple most strongly due to a $r^{-3}$ dependence while higher order multipoles such as quadrupole and octupole couple weakly. The dotted ($d_p = 30$ nm) and dashed ($d_p = 50$ nm) curves and the solid curve in the inset ($d_p = 100$ nm) in Fig. 7 have a lower magnitude in comparison to the solid curve as a result. In fig. 7b, the solid curve for $d_p = 10$ nm shows a prominent radiant peak for $\lambda \approx 610$ nm with $\Re(B) \approx -3000$. For larger particles such as the ones with $d_p =$ 30 and 50 nm, peak values of $\Re(B)$ are greatly diminished due to reduced coupling. Hence, the dashed curves in Fig. 7b, have a peak at $\sim -900$ and the dotted curve at $\sim -100$. The corresponding $A$ values do not vary proportionately as can be seen from Fig. 6. Hence, linear approximation is appropriate only for large $d_p$ values. This is not surprising since for a given $f$, increasing $d_p$ results in reducing particle number density, and consequently interparticle coupling. The linear approximation is also applicable to well-separated systems since they have small values of $B$ in comparison to a random system. A maximum in $\Re(B)$ also occurs upon increasing $d_p$ while keeping $f$ and $\lambda$ constant leading to a size resonance. Thus, $\varepsilon_{eff}$ exhibits resonances as a function of all system parameters.

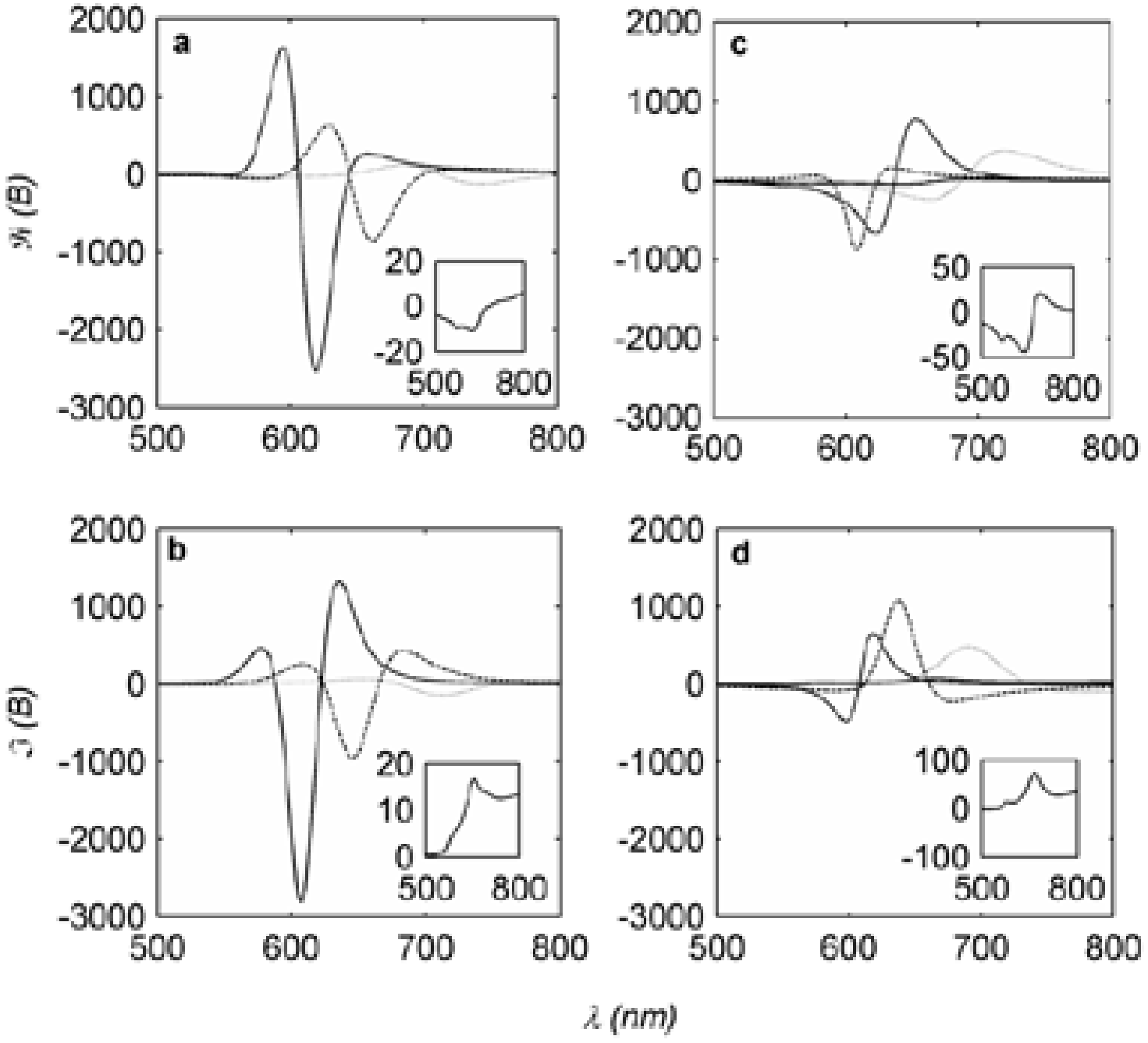

**Figure 7**: Quadratic coefficient $B$ calculated for random (a, b) and well-separated random (c, d) composites in an $\varepsilon_m = 7$ matrix containing Ag NPs with diameters $d_p = 10$ (solid), 30 (dashed), 50 (dotted) and 100 nm (inset). Here, $k^* a \approx \frac{0.45 d_p \sqrt{\varepsilon_p}}{\lambda} < 1$ is less than one only for the solid curve that is given by $\frac{3}{4}(\beta + 4)\beta^2$ for (a) and (b); and $3\beta^2$ for (c) and (d). Weak coupling in well-separated random systems leads to a smaller $B$ in comparison to random systems.

## C. Fano Resonance

High conductivity metals are typically described by adding to the Drude model a number of Lorentz oscillators that capture effects of lattice polarizability and inter-band and intra-band electron transitions [55]. The Lorentz model is based on a damped harmonic oscillator with finite mass. The Drude model, however, does not include the harmonic force and hence is able to model free electrons well. The Lorentz model predicts a symmetric profile for the intensity vs. frequency curve for systems with a small damping. Interestingly, $\beta$ that is based on a Drude model for $\varepsilon_p$ has a lineshape of a Lorentz oscillator. Hence, a single plasmonic particle is a Lorentz oscillator. In the limit as $f \to 0$, the susceptibility of the effective medium is given by $\chi_{eff} = \varepsilon_{eff} / \varepsilon_m - 1 = 3\beta f + O(f^2)$. Hence, the effective medium is also a Lorentz oscillator in the limit as $f \to 0$. However, the coupling between the particle and the effective medium becomes stronger, i.e., for large $f$ or small $\kappa$, $\varepsilon_{eff}$ can be expected to possess the characteristics of a coupled oscillator system which deviates from the Lorentzian symmetric lineshape. It is well-known that an unusual lineshape that is asymmetric about the extremum, known as Fano resonance, is observed in resonant coupled oscillator systems such as plasmonic nanostructures [56-59].

Within the framework of the EMT presented here, for a given $f$, $\kappa$ represents the extent of coupling between the particle and the effective medium. Consequently, $\varepsilon_{eff} - \lambda$ curves show unusual resonance shapes for relatively small values of $\kappa$. Fig. 8 shows a plot of $\varepsilon_{eff}$ for Ag NP composite in an $\varepsilon_m = 7$ matrix for $f = 5\%$. The scalar EMT was used in the calculations. Hence, the results are representative of those for small particles. For random systems, $\kappa$ has two bounds: an upper bound given by $\kappa = f^{-1/3}$ for a system with well-separated particles and a lower one given by $\kappa = 1$ representing a locally dense composite, which we refer to as the Bruggeman limit. Close to the upper bound, the particle-effective medium coupling is relatively weak resulting in a symmetric Lorentzian $\varepsilon_{eff}$ even for relatively large $f$. This can be seen in the shape of the solid curve in Fig. 8. The dashed curve in Fig. 9 is the locus for the upper bound in the $\kappa$-$f$ space. Black circles in Fig. 9 denote the locations at which the shape of $\varepsilon_{eff}$-$\lambda$ curve is Lorentzian. As $\kappa$ is reduced, the shapes of the curves become distorted due to increased coupling. The resonance location shifts to larger values of $\lambda$ and the resonant peak becomes broader. $\Re(\varepsilon_{eff} / \varepsilon_m)$ has an asymmetric shape for intermediate values of $\kappa$ as shown in dashed curves in Fig. 8. The $\kappa$-$f$ space locations for which an asymmetric response is predicted are shown in Fig. 9 as grey circles. For $\kappa$ values that are close to unity, $\Re(\varepsilon_{eff} / \varepsilon_m)$ curve becomes very broad as can be seen from the dotted lines in Fig. 8. Unfilled circles in Fig. 9 represent this type of response. The trends in Figs. 8 and 9 are also present in composites with a finitely large *ka*.

Fano resonance can also be understood as the interference between the absorbing and radiating modes in plasmonic structures [56-57]. Our analysis in the limit as $f$ ® 0 suggests that the interference manifests through the Taylor coefficients of $e_{eff}$ . A negative extreme in the imaginary quadratic coefficient Á$(B)$ (Fig. 7b and d), for example, represents a coupled "radiating mode" that is present in conjunction with the "absorbing mode" of a single particle given by a positive Á$(A)$ (Fig. 6b).

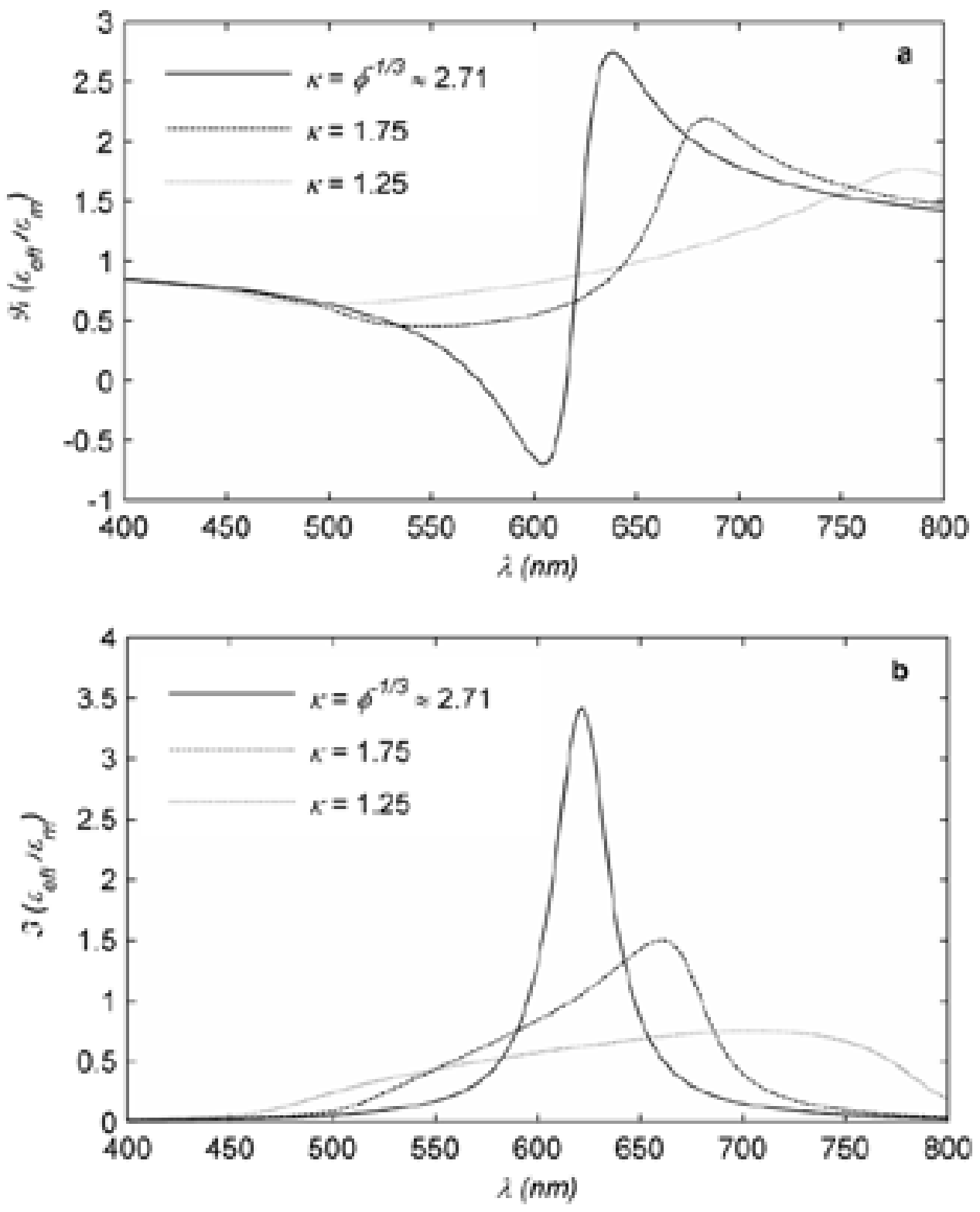

**Figure 8**: Effect of $k$ on $e_{eff}$ for $f = 5\%$ . Small values of $k$ lead to a stronger coupling that distorts the Lorentzian shape of $e_{eff}$ even for relatively small values of $f$ such as 5%. Calculations were performed with the scalar EMT.

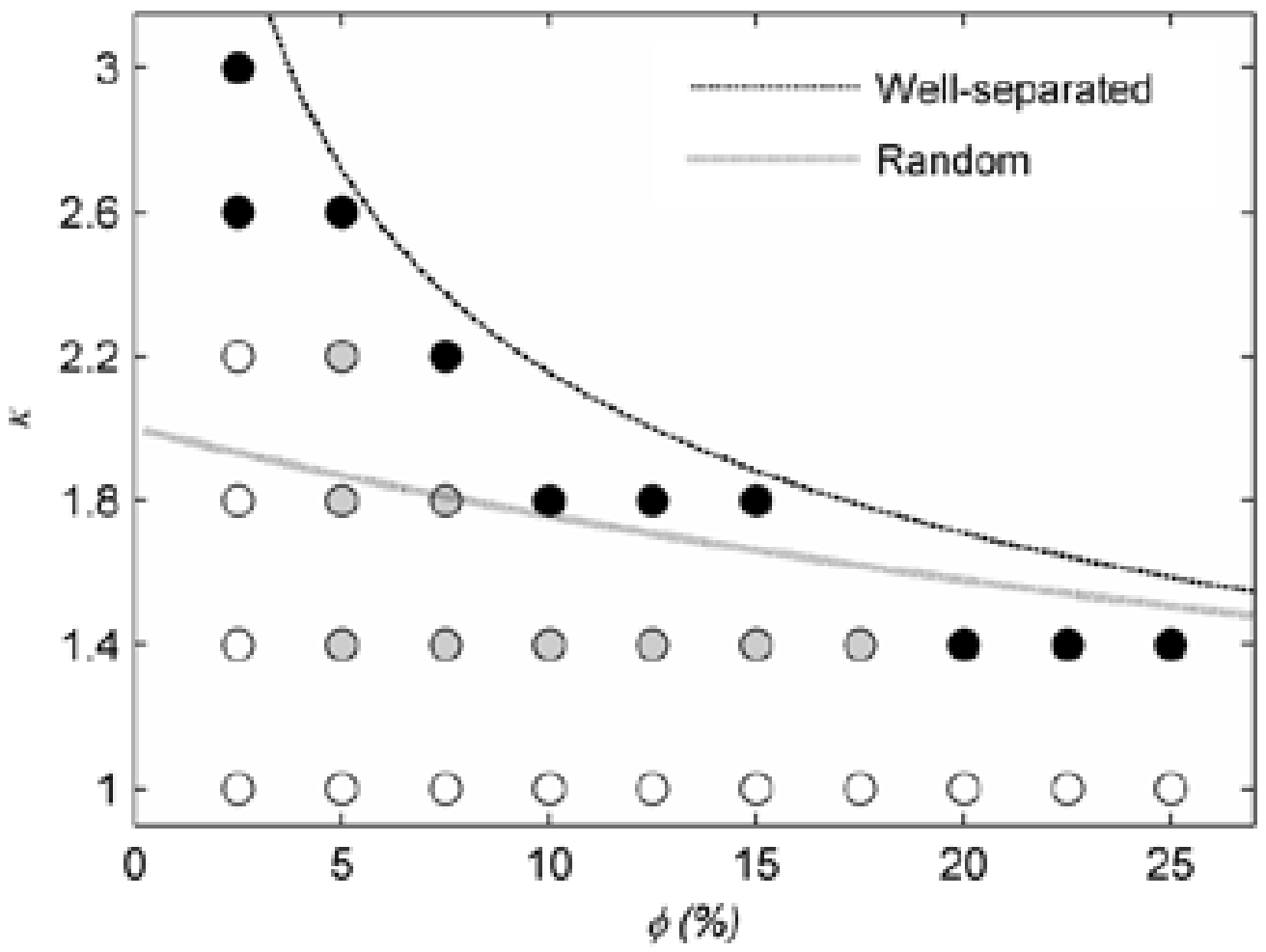

**Figure 9**: Characterization of the dielectric response of a random plasmonic composite in the $k$-$f$ space. The dashed line represents the upper bound for a random composite with well-separated particles and the grey dotted line represents a random hard sphere composite. Locations of Lorentzian and Fano responses are shown in black and grey circles respectively. Unfilled circles denote locations in which broad lineshapes are observed.

## 5. Conclusions

Plasmonic nanoparticles undergo an electric resonance when their electric dipole polarizability $b$ is a large positive imaginary number. As a result, the particles absorb intensely. For a relatively large volume fraction $f$ and/or when particle size is comparable to that of the exciting wavelength, interparticle coupling becomes important. Consequently, the effective permittivity $e_{eff}$ can no longer be treated as the sum of the polarizabilities of the individual particles and, hence does not vary linearly with $f$. Hence, quadratic and higher order effects in $f$ on the polarizability have to be determined to obtain accurate predictions of $e_{eff}$.

In this work, we have developed an EMT to account for such nonlinear effects on the effective permittivity of dense random dispersions of equi-sized spheres of high optical-conductivity metals such as Ag, Au and Cu. The EMT is based on the idea that the region surrounding a given particle in a composite can be modeled as an effective continuum that begins after a distance $R=\kappa a$ from the centre of the particle ($\kappa>1$). Within this framework, $\kappa$ is interpreted as a microstructure parameter that correlates with the static structure factor of the composite. For a homogenous random composite, $\kappa$ is bounded such that $1<k<f^{-1/3}$. The upper bound corresponds to a well-separated random system that can be modeled as a Maxwell-Garnett composite. The lower bound corresponds to Bruggeman's mixing rule which, as in the case of MGT, is based on the electrostatic approximation [27]. In general, a random hard-sphere microstructure would have a $\kappa$ value that lies in between these two bounds, which, in principle can be determined from the knowledge of the radial distribution function. Hence $k$ is a *physical* rather than an adjustable fitting parameter. Conversely, if $k$ were to be determined by fitting spectroscopic data to the EMT predictions, it can be used to better understand the internal microstructure of the composite.

The scalar (electrostatic) approximation is valid only for particles much smaller than the exciting wavelength. The EMT presented here takes into account both the microstructure and finite size effects in a self-consistent fashion. Specifically, two scenarios were examined, one in which $ka$ ® $0$ in which the conditionally averaged electric field can be obtained by the solution of the Laplace equation for the electrostatic potential (Scalar EMT) and a more general case for finitely large spheres for which a solution of vector Helmholtz equation for **E** is required (Vector EMT).

Resonance conditions for individual particles were found to depend on $k$ and $f$. In the limit as $ka$ ® $0$, the scaling of the particle polarizability at resonance with $f$ depends on the microstructure. A well-separated random composite has an $e_{eff}$ resonance when $b$ » $1/f$. In contrast, for random hard-sphere composite the resonance condition is given by $b$ » $2/\sqrt{f}$. Hence, for a given $f$, the resonance peak is more red-shifted for random systems. For finite sized spheres, the vector EMT problem was solved numerically to obtain the a quadratic approximation for $e_{eff}$ as a function of $f$. As the particle size is increased, $e_{eff}$ vs. $l$ curve exhibit multiple peaks corresponding to quadrupolar, octupolar, and higher order resonances in addition to the dipolar resonance. Size effect on $e_{eff}$ becomes significant when $\left|b^2\left(k^*a\right)^3\right|$ is $O(1)$. Hence, for composites consisting of high conductivity metals such as Ag in a medium with large refractive index in the visible range such as $TiO_2$, ZnO and Si, size effects could manifest even for particle diameters of a few 10s of nm.

Particle-effective medium coupling gives rise to a non-Lorentzian resonance behavior in $e_{eff}$. In order to characterize the resonant optical response in the $k - f$ space, a "phase diagram" was constructed. Three regions were identified based on the lineshape of Á($e_{eff}$) :(i). Lorentzian (symmetric peak), (ii). Fano (distinctly asymmetric peak), and (iii). Bruggeman (broad). For an Ag NPs in a high refractive index medium, Fano resonance region is enveloped by the Lorentzian (large $k$ or large $f$) and Bruggeman (small $k$ or small $f$) regions. Overall, the predictions of the EMT are in qualitative agreement with experimental trends observed for plasmonic composites [13]. This work motivates experimental investigations to quantify the effect of volume fraction on the optical response of plasmonic nanocomposites with well-characterized microstructures. EMT presented here can be extended to describe non-linear response of polydisperse and multiple species systems by adapting methodologies described elsewhere [39, 50, 60].

**Acknowledgment**

RS and SW would like to thank National Science Foundation for financial support through Grants No. CMMI 0757589 and CBET 1049454.

## Appendix A: $c_n$ and $d_n$

Mie coefficients, $c_n$ and $d_n$, for **E** inside a particle in a layered sphere geometry shown in Fig. 1 are discussed in §3. Hightower and Richardson showed that they can be calculated in the following fashion [44].

$$c_n = \left[\frac{k_p \psi_n(k_m a) - B_n k_p \chi_n(k_m a)}{k_m \psi_n(k_p a)}\right]\left[\frac{k_m \psi_n(k_{eff} R) - b_n k_m \xi_n(k_{eff} R)}{\psi_n(k_m R) - B_n \chi_n(k_m R)}\right] \tag{A1a}$$

and

$$d_n = \left[\frac{\psi_n(k_m a) - A_n \chi_n(k_m a)}{\psi_n(k_p a)}\right]\left[\frac{\psi_n(k_{eff} R) - a_n \xi_n(k_{eff} R)}{\psi_n(k_m R) - A_n \chi_n(k_m R)}\right]. \tag{A1b}$$

Here, $k_n, n = p,m,eff$ denotes the wavenumber in the particle, medium and effective medium respectively. Particle radius is $a$ and $R$ is the shell radius as shown in Fig. 1. Riccati-Bessel functions $\psi_n$, $\chi_n$ and $\xi_n$ are defined as:

$$\psi_n(z) \equiv z j_n(z), \tag{A2a}$$

$$\chi_n(z) \equiv -z y_n(z) \tag{A2b}$$

and

$$\xi_n(z) \equiv z h_n^{(1)}(z). \tag{A2c}$$

Here, $j_n$, $y_n$ and $h_n^{(1)}$, respectively, are the regular, singular and outward propagating spherical Bessel functions. Note that spherical Hankel functions of the first kind are defined as: $h_n^{(1)} \equiv j_n + i y_n$ [45]. The coefficients $A_n$ and $B_n$ have the following form:

$$A_n = \frac{k_m \psi_n(k_m a) \psi_n'(k_p a) - \psi_n'(k_m a) \psi_n(k_p a)}{k_m \chi_n(k_m a) \psi_n'(k_p a) - \chi_n'(k_m a) \psi_n(k_p a)} \tag{A3a}$$

and

$$B_n = \frac{k_m \psi_n'(k_m a)\psi_n(k_p a) - \psi_n(k_m a)\psi_n'(k_p a)}{k_m \chi_n(k_m a)\psi_n'(k_p a) - \chi_n'(k_m a)\psi_n(k_p a)}. \tag{A3b}$$

The primes in Eqs. A3 denote a differentiation. The coefficients $a_n$ and $b_n$ in Eqs. A1 have the following form:

$$a_n = \frac{\psi_n(k_{eff}R)\left[\psi_n'(k_m R) - A_n\chi_n'(k_m R)\right] - k_m\psi_n'(k_{eff}R)\left[\psi_n(k_m R) - A_n\chi_n(k_m R)\right]}{\xi_n(k_{eff}R)\left[\psi_n'(k_m R) - A_n\chi_n'(k_m R)\right] - k_m\xi_n'(k_{eff}R)\left[\psi_n(k_m R) - A_n\chi_n(k_m R)\right]} \tag{A4a}$$

and

$$b_n = \frac{k_m\psi_n(k_{eff}R)\left[\psi_n'(k_m R) - B_n\chi_n'(k_m R)\right] - \psi_n'(k_{eff}R)\left[\psi_n(k_m R) - B_n\chi_n(k_m R)\right]}{k_m\xi_n(k_{eff}R)\left[\psi_n'(k_m R) - B_n\chi_n'(k_m R)\right] - k_m\xi_n'(k_{eff}R)\left[\psi_n(k_m R) - B_n\chi_n(k_m R)\right]}. \tag{A4b}$$